\documentclass[twocolumn,english,prl, showpacs, superscriptaddress]{revtex4-1}
\usepackage[T1]{fontenc}
\usepackage[utf8]{inputenc}
\usepackage{color}
\usepackage{babel}
\usepackage{amsmath}
\usepackage{amssymb}
\usepackage{graphicx}
\usepackage[unicode=true,pdfusetitle,bookmarks=true,bookmarksnumbered=false,bookmarksopen=false,breaklinks=false,pdfborder={0 0 1},backref=false,colorlinks=false]{hyperref}

\makeatletter

\hypersetup{
    colorlinks=true,
    linkcolor=red,
    citecolor=blue,
    filecolor=blue,      
    urlcolor=blue,
}
\makeatother

\begin{document}

\title{\textcolor{black}{Negative temperature is cool for cooling}}

\author{Gabriella G. Damas}
\address{Instituto de Física, Universidade Federal de Goiás, 74.001-970, Goiânia - GO, Brazil}
\author{Rogério J. de Assis}
\address{Instituto de Física, Universidade Federal de Goiás, 74.001-970, Goiânia - GO, Brazil}
\address{Departamento de Física, Universidade Federal de São Carlos, 13.565-905, São Carlos - SP, Brazil}
\author{Norton G. de Almeida}
\address{Instituto de Física, Universidade Federal de Goiás, 74.001-970, Goiânia - GO, Brazil}


\begin{abstract}

In this work, we study an autonomous refrigerator composed of three qubits {[}Phys. Rev. Lett. 105, 130401 (2010){]} operating with one of the reservoirs at negative temperatures, which has the purpose of cooling one of the qubits. We find the values of the lowest possible temperature that the qubit of interest reaches when fixing the relevant parameters, and we also study the limit for cooling the qubit arbitrarily close to absolute zero. We thus proceed to a comparative study showing that reservoirs at effective negative temperatures are more powerful than those at positive temperatures for cooling the qubit of interest. 

\end{abstract}

\maketitle

\section{Introduction}

In a 2010 work, Linden et al. (Ref. \citep{Linden2010}) studied three models for cooling qubits or arbitrary quantum systems. The first model uses only qubits, the second one a qubit and a qutrit with nearest-neighbor interactions, and the third one a single qutrit. The model using only qubits is a self-contained or autonomous refrigerator (AR) with three interacting qubits, one of them being the object to be cooled. As shown in Ref. \citep{Linden2010}, under the condition of perfect insulation of the qubit to be refrigerated, this model presents no fundamental cooling limit, and thus, depending on the parameters used, the qubit of interest can be arbitrarily cooled towards absolute zero.

In this work, we study the AR consisting only of qubits as proposed by Linden et al. in Ref. \citep{Linden2010}, putting it to work with one of the reservoirs at negative temperature. Negative absolute temperatures were initially considered in 1951 when Purcell first produced spin states with population inversion \citep{Purcell1951}. In 1956, Ramsey theoretically studied these states, treating them as thermodynamic equilibrium states \citep{Ramsey1956}. More than 60 years after the experiment performed by Purcell, other experiments involving negative temperatures followed \citep{Carr2013,Braun2013}, which drew attention to this topic. While some authors defend the existence of negative thermodynamic temperatures, some works associate them with non-equilibrium states and thus refer to them as effective (or apparent) negative temperatures, as opposed to equilibrium states at positive thermodynamic temperatures. \citep{Dunkel2013,Sokolov2013,Hilbert2014,Campisi2015,Cerino2015,Frenkel2015,Swendsen2015,Poulter2016,Swendsen2016,Abraham2017,Hama2018,Struchtrup2018}. As a general property, experimental realizations of reservoirs with effective negative temperatures, such as the ones we will consider in this work, requires population inversion \citep{Carr2013,Braun2013,Assis2019,Taysa2020}. As we will show, the qubit-based model AR has advantages when operating with one of the reservoirs at negative temperature compared to operating only with conventional reservoirs at positive temperatures. 

This work is organized as follows: In Section II, we present the model for the study of AR considering bosonic reservoirs, for which the temperature is always positive, and fermionic reservoirs, which can reach negative temperatures. In Section III, we present our results, showing that the use of a fermionic reservoir with negative temperatures has an advantage over the use of only bosonic reservoirs. Finally, in Section IV, we present our conclusions about the comparative study of the AR involving positive and negative temperatures.

\section{Model}

\begin{figure}
\centering{}\includegraphics{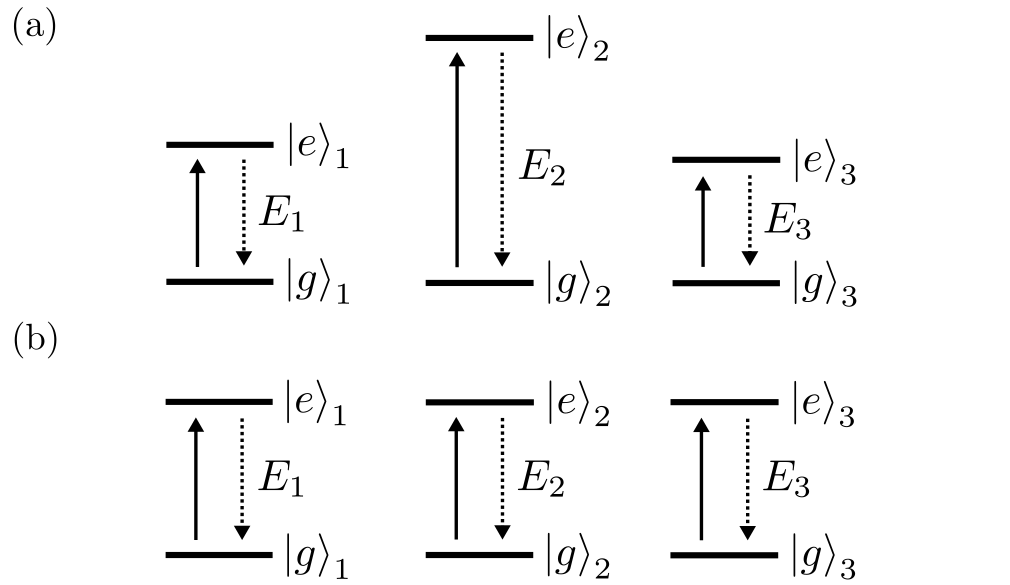}\caption{\label{fig: 1} Schematic representation of possible level configuration
for the qubit-based AR cooler. (a) Level configurations used
in Ref. \citep{Linden2010} for positive temperatures, where the condition
$E_{3}=E_{2}-E_{1}$ is required. (b) A possible level configuration
with $E_{3}=E_{2}=E_{1}$ for negative temperatures, where the condition
$E_{3}=E_{2}-E_{1}$ is no longer needed. In (a) and (b),
$\left|g\right\rangle _{k}$ and $\left|e\right\rangle _{k}$ are
the ground and excited states of the $k$th qubit, respectively.}
\end{figure}

The AR considered here consists of three qubits, each in contact with a different reservoir \citep{Linden2010}. The first qubit, which is the qubit to be cooled, is put in contact with a cold reservoir at temperature $T_{c}$. The second qubit, analogously to the classical refrigerator, plays the role of the spiral and is put in contact with a reservoir at \textquotedbl room\textquotedbl{} temperature $T_{r}$, while the third qubit works as the engine staying in contact with a hot reservoir at temperature $T_{h}$. For the AR to work properly, considering only reservoirs at positive temperatures, the relations $T_{c}<T_{r}<T_{h}$ and $E_{3}=E_{2}-E_{1}$ - see Fig. \ref{fig: 1}(a), must be satisfied \citep{Linden2010}.

In the weak coupling limit and the Markovian regime, the dynamics governing the AR is dictated by the master equation
\begin{multline}
\frac{\partial\rho}{\partial t}=-i\left[H_{0}+H_{int},\rho\right]+\\
\frac{1}{2}\sum_{k=1}^{3}\Gamma_{k}^{\downarrow}\left(2\sigma_{-,k}\rho\sigma_{+,k}-\left\{ \sigma_{+,k}\sigma_{-,k},\rho\right\} \right)+\\
\frac{1}{2}\sum_{k=1}^{3}\Gamma_{k}^{\uparrow}\left(2\sigma_{+,k}\rho\sigma_{-,k}-\left\{ \sigma_{-,k}\sigma_{+,k},\rho\right\} \right),\label{eq: 1}
\end{multline}
where $H_{0}$ and $H_{int}$ are respectively the free qubits Hamiltonian and the three-body interaction Hamiltonian, given by 
\begin{equation}
H_{0}=\frac{1}{2}E_{1}\sigma_{z,1}+\frac{1}{2}E_{2}\sigma_{z,2}+\frac{1}{2}E_{3}\mathbb{\sigma}_{z,3}
\end{equation}
and 
\begin{equation}
H_{int}=g\left(\sigma_{-,1}\sigma_{+,2}\sigma_{-,3}+\sigma_{+,1}\sigma_{-,2}\sigma_{+,3}\right).
\end{equation}
In the above equations, $\sigma_{-,k}$, $\sigma_{+,k}$, and $\sigma_{z,k}$ \textbf{($k=1,2,3$)} are respectively the lowering, raising, and z Pauli operators for the $k\text{th}$ qubit; $E_{1}$, $E_{2}$, and $E_{3}$ are the energy gaps for each qubit, and $g$ is the coupling constant. Also, we let $\Gamma_{k}^{\downarrow}=\gamma_{k}\left(1\pm n_{\alpha_{k}}\right)$, where the signal "+" ("-") is for a bosonic (fermionic) reservoir, and $\Gamma_{k}^{\uparrow}=\gamma_{k}n_{\alpha_{k}}$ ($\alpha_{1}=c$, $\alpha_{2}=r$, and $\alpha_{3}=h$), with $\gamma_{k}$ being the dissipation rate for the $k$th qubit and $n_{\alpha_{k}}$ being the average excitation number of the reservoir at temperature $T_{{\alpha_{k}}}$. Note that Eq. \eqref{eq: 1} describes the dynamics dictated by either bosonic \citep{Carmichael1999, Breuer2007} or fermionic \citep{Artacho1993,Ghosh2012} thermal reservoirs. For bosonic thermal reservoirs $n_{\alpha_{k}}=1/\left(\text{e}^{E_{k}/T_{\alpha_{k}}}-1\right)$, while for fermionic reservoirs $n_{\alpha_{k}}=1/\left(e^{E_{k}/T_{\alpha_{k}}}+1\right)$. As said before, negative temperatures are characterized by population inversion, which occurs when $n_{\alpha_{k}}>1/2$, and it is only possible for fermionic reservoirs. To find the final temperature of qubit 1, we numerically calculate the asymptotic equilibrium state of Eq. \eqref{eq: 1}. We perform these numerical calculations using the quantum optics toolbox \citep{Johansson2012,Johansson2013}.

\section{Results}

\begin{figure}
\centering{}\includegraphics{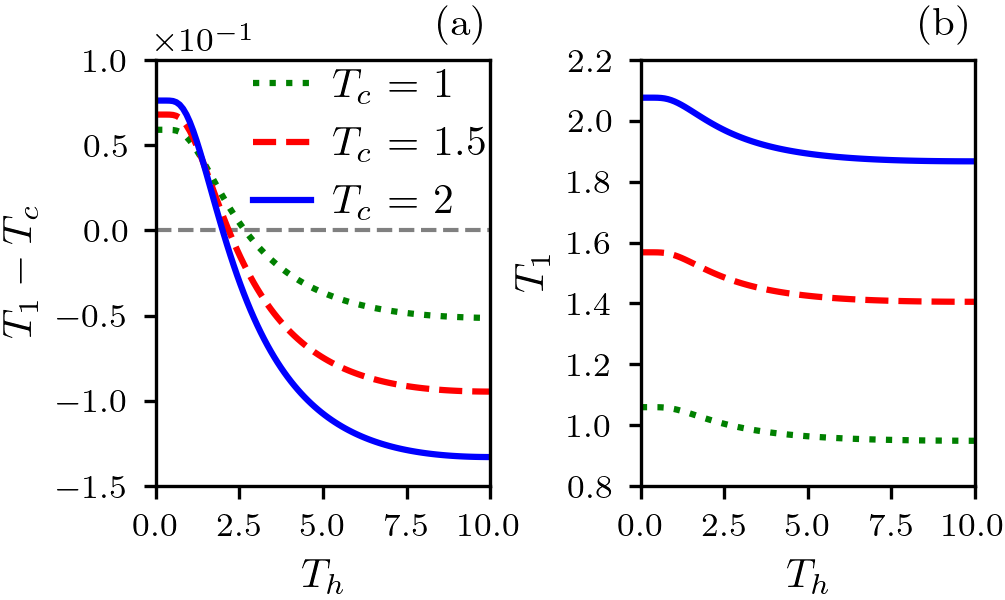}\caption{\label{fig: 2} (a) $T_{1}-T_{c}$ versus $T_{h}>0$ for three different
values of $T_{c}$: $T_{c}=1$ (dot green line), $T_{c}=1.5$ (dash
red line) and $T_{c}=2$ (solid blue line). Refrigeration occurs for
$T_{1}-T_{c}<0$. (b) $T_{1}$ behavior for those three initial conditions
given in (a). Note that the lowest temperature of the qubit $T_{1}$
does not decrease any further, no matter how high $T_{h}$ becomes.
The lowest values reached by $T_{1}$ are $T_{1}=0.9486$ (when $T_{c}=1$),
$T_{1}=1.4054$ (when $T_{c}=1.5$) and $T_{1}=1.867$ (when $T_{c}=2$).}
\end{figure}

To compare the AR operation at positive temperature with its operation at negative temperature, we must consider the same parameters that characterize the AR, such as coupling constant, energy gaps and dissipation rates. For clarity, let us start detailing the positive temperature case by fixing the parameters used in Ref. \citep{Linden2010}. Thus, we fixed the energies gaps $E_{1}=1$, $E_{2}=5$, and $E_{3}=4$, which is a special case of the more general relation $E_{3}=E_{2}-E_{1}$ to AR work properly, and the temperatures $T_{c}=1,1.5,2$, and $T_{r}=2$. Finally, we fixed the dissipation rates $\gamma_{1}=\gamma_{2}=\gamma_{3}=1$ and let $T_{h}$ vary from $1$ to $10$. Cooling is achieved when $T_{1}-T_{c}<0$, which means that the temperature $T_{1}$ of the qubit $1$ is lower than the temperature $T_{c}$ of the cold reservoir -
See Fig. \ref{fig: 2}(a). In Fig. \ref{fig: 2}(b) we show how the temperature $T_{1}$ drops by increasing $T_{h}$. Note from Fig. \ref{fig: 2}(b)
that the temperature of qubit 1 stabilizes at a constant value no matter how high $T_{c}$ is. We numerically found that the corresponding lowest temperatures reached by the qubit 1 are $T_{1}=0.9486$ (when $T_{c}=1$), $T_{1}=1.4054$ (when $T_{c}=1.5$), and $T_{1}=1.867$ (when $T_{c}=2$). These lowest possible temperatures, when fixing the set of parameters, are dictated by the population of the ground state $p_{g,1}$, which is given by $p_{g,1}=1/\left(1+e^{-E_{1}/T_{1}}\right)$. Solving for temperature, we get 
\begin{equation}
T_{1}=\frac{E_{1}}{\ln\left(\frac{p_{g,1}}{1-p_{g,1}}\right)}.\label{eq: 4}
\end{equation}
From this equation, we see that $T_{1}$ can reach any value since $0.5<p_{g,1}<1.0$. However, due to imperfect insulation, the heat flow to qubit 1 prevents it from reaching temperatures arbitrarily close to absolute zero. In fact, in our numerical calculations, we found that if we start at low temperatures as $T_{c}<0.48$, then $T_{1}-T_{c}>0$, meaning that the refrigerator no longer works. As we shall see, using the same parameters but considering one of the reservoirs at negative temperatures, we can cool the qubit 1 below $T_{c}=0.48$.

Next, we use the perfect insulation condition for the temperature of qubit 1 to get as close to absolute zero as possible \citep{Linden2010}. This condition is obtained by isolating qubit 1 from its reservoir, thus letting $\gamma_{1}\rightarrow0$. This condition allows us to reach lower temperatures for qubit 1 than that shown in Fig. \ref{fig: 2}(b) and obtain the following analytical expression for the cooling temperature \citep{Linden2010}
\begin{equation}
T_{1}=\frac{T_{c}}{1+\frac{E_{3}}{E_{1}}\left(1-\frac{T_{c}}{T_{h}}\right)}.\label{eq: 5}
\end{equation}
The above formula tell us that, under the condition of perfect insulation, it is possible to cool down to absolute zero, provided that $E_{3}/E_{1}\rightarrow\infty$.

Now we turn our attention to negative temperatures. In this case, we must indicate which reservoir will be at negative temperatures. Knowing that the engine is modeled by qubit 3, which is in contact with a reservoir at hot temperature $T_{h}$, it is natural to replace the reservoir of qubit 3 by a reservoir at $T_{h}<0$ since bodies at negative temperatures are known to be hotter than any body at positive temperature as it scales, from cold to hot, according to $+0\text{ K},...,+300\text{ K},...,+\infty \text{ K}$, $-\infty \text{ K},...,-300\text{ K},...,-0\text{ K}$. In view of that, one might think that instead of using negative temperatures it would suffice to use arbitrarily hot reservoirs at positive temperatures. However, we saw from Fig. \ref{fig: 2}(b) that this is not the case: no matter how high $T_{h}$ is, there is a limit achieved by $T_{1}$. Remarkably, we will show that under the same conditions, i.e., the same set of parameters characterizing the AR, the lowest possible temperature of qubit 1 cooled using positive temperatures
can be pushed even lower using a hot reservoir at negative temperatures. 

\begin{figure}
\centering{}\includegraphics{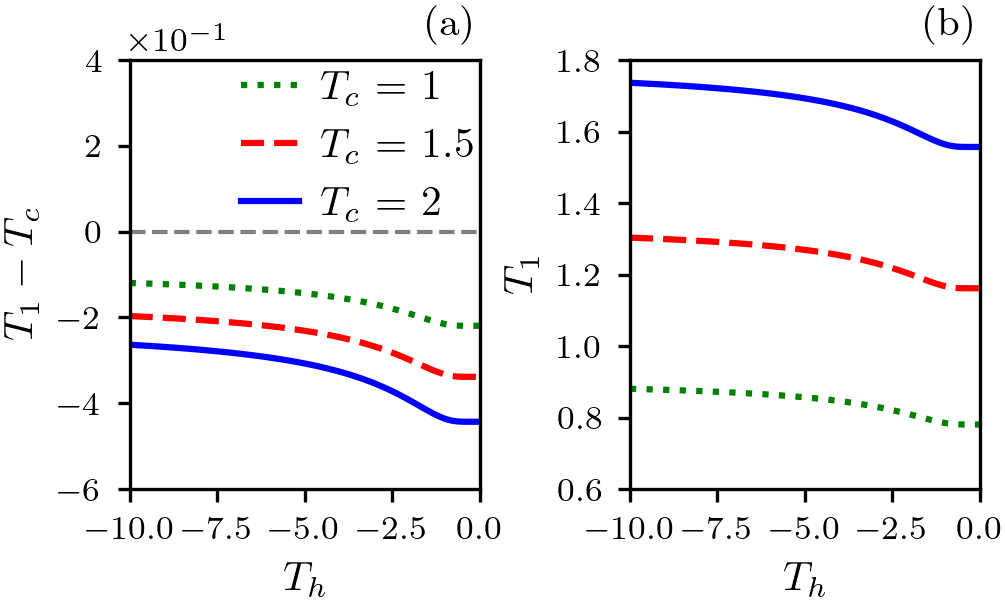}\caption{\label{fig: 3} (a) $T_{1}-T_{c}$ versus $T_{h}<0$ using the same
three values for $T_{c}$ as before: $T_{c}=1$ (dot green line),
$T_{c}=1.5$ (dash red line) and $T_{c}=2$ (solid blue line). Refrigeration
occurs for $T_{1}-T_{c}<0$. Note that the higher the temperature
$T_{h}$ the greater the cooling of qubit 1, as before, but now the
steady state of qubit 1 is cooled for all negative temperatures. (b)
$T_{1}$ versus $T_{h}$ for those three initial conditions given
in (a). Again, the lowest temperature of the qubit $T_{1}$ does not
decrease any further by increasing $T_{h}$. However, this time the
lower temperatures are closer to absolute zero than when using positive
temperatures. The corresponding lowest temperatures are $T_{1}=0.7805$
(when $T_{c}=1$ ), $T_{1}=1.1615$ (when $T_{c}=1.5$) and $T_{1}=1.5568$
(when $T_{c}=2$).}
\end{figure}

As said before, to compare the cooling process at either negative or positive temperatures, we put the AR to work with the same parameters. The steady state is found using the Eq. \eqref{eq: 1}, but now considering a fermionic reservoir with $n_{3}>0.5$ ($T_{h}<0$). In Fig. \ref{fig: 3}(a), we show the difference $T_{1}-T_{c}$ as a function of $T_{h}<0$. Interestingly, if we compare with Fig. \ref{fig: 2}(a) for $T_{h}>0$, we see that at negative temperatures the steady-states are always cooled, no matter the initial temperature we choose for $T_{h}<0$. Also, remembering that for $T_{h}<0$ the higher temperatures ranges according to $\left(-\infty,0^{-}\right)$ in Fig. \ref{fig: 3}(b) we see that the higher the temperature $T_{h}$ the lower is the temperature of qubit 1.

\begin{figure}
\includegraphics{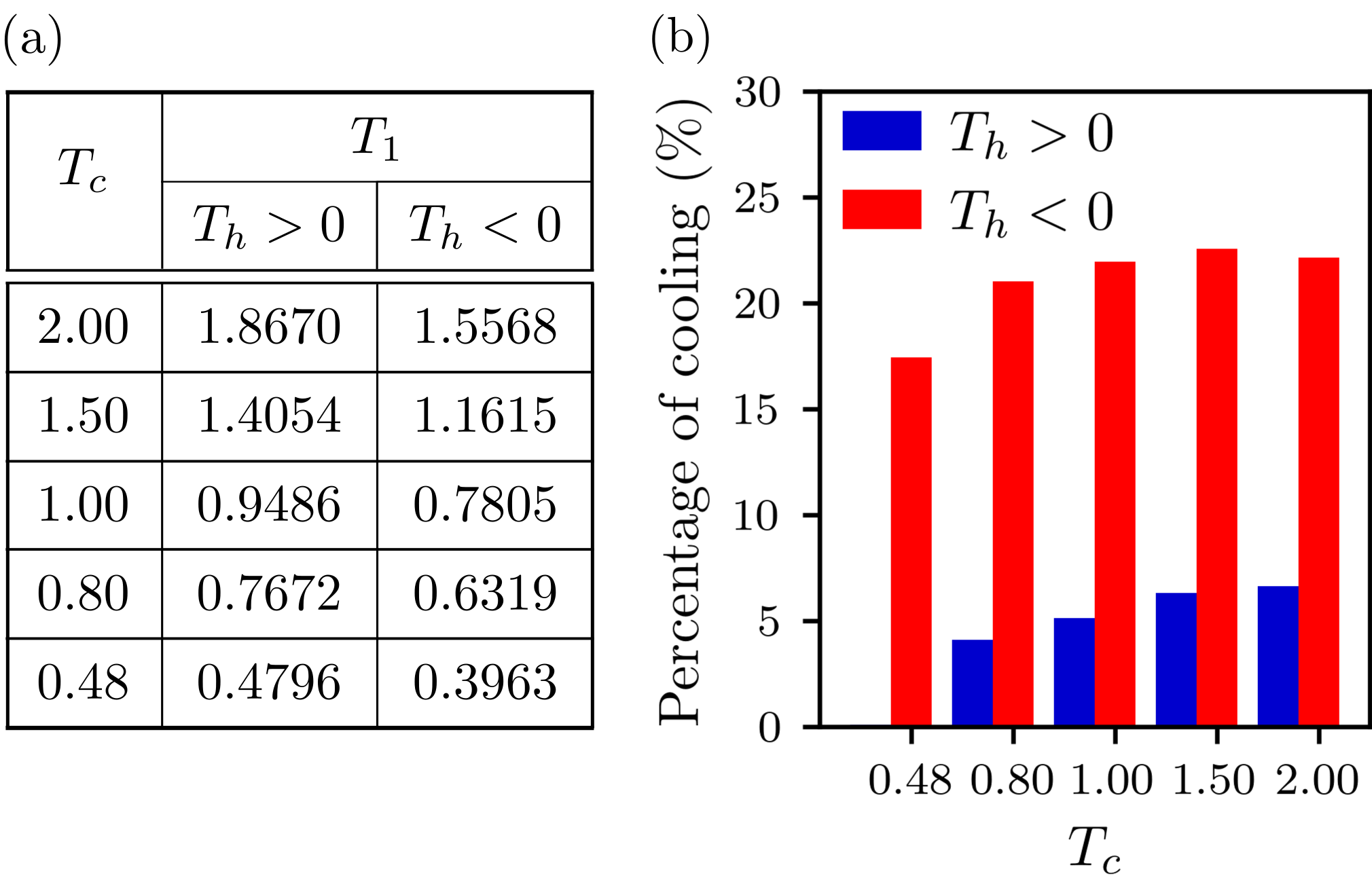} \caption{\label{fig: 4}(a) Lowest values for $T_{1}$ obtained when either $T_{h}>0$
or $T_{h}<0$ for some values of $T_c$ (b) Percentage of cooling using both positive and negative
temperatures for the pumping heat at $T_{h}$ and taking $T_{c}$
as reference. According to our numerical calculations, AR stops working if $T_{c}< 0.48$ ($T_{c}< 0.0275$) when $T_{h}>0$  ($T_{h}<0$).}

\end{figure}

Note that the temperature reached by qubit 1 is yet given by equation Eq. \eqref{eq: 4}. Again, imperfect insulation of qubit 1 prevents it from reaching temperatures arbitrarily close to absolute zero, as also shown in Figs. \ref{fig: 3}(a) and \ref{fig: 3}(b), similarly to what happens in the case of $T_{h}>0$. To the set of parameters we
are using, these values are $T_{1}=0.7805$ (when $T_{c}=1$), $T_{1}=1.1615$
(when $T_{c}=1.5$), and $T_{1}=1.5568$ (when $T_{c}=2$). Interestingly, note that
the lowest temperatures $T_{1}$ in the case $T_{h}<0$ are lower
than the temperatures $T_{1}$ corresponding to the case $T_{h}>0$.
That is true for all reference temperatures $T_{c}$ according to
our numerical calculation and as can be seen in few examples shown in the table of Fig. \ref{fig: 4}(a). Taking $T_{c}$ as the reference temperature, we
can calculate the percentage decrease of $T_{1}$ in relation
to $T_{c}$ for both negative and positive temperatures. In Fig. \ref{fig: 4}(b)
we show the corresponding percentages for both $T_{h}>0$ (blue) and
$T_{h}<0$ (red).

Fig. \ref{fig: 4} makes it clear that negative temperatures are more effective
for cooling a qubit, as it drives qubit 1 to lower temperatures by
a large percentage. According to our numerical calculations, for this
set of parameters, which is the same as used for positive temperatures,
when the reference temperature $T_{c}$ is as low as $T_{c}=0.48$
the AR stops cooling if $T_{h}>0$. However, at $T_{h}<0$
the AR continues to cool down until qubit 1 reaches the limit
of $T_{1}=0.0275$, that is, an order of magnitude lower.

We can also ask what happens when we impose the condition for perfect
insulation, $\gamma_{1}\rightarrow0$. In this case, we can demonstrate
that Eq. \eqref{eq: 5} remains valid. Alternatively, we can write
Eq. \eqref{eq: 5} as

\begin{equation}
T_{1}=\frac{T_{C}}{1+\frac{E_{3}}{E_{1}}(1+\frac{T_{C}}{\left|T_{h}\right|})}.\label{eq: 6}
\end{equation}
Eq. \eqref{eq: 6} tells us that, as in the case of positive temperatures,
it is possible to cool a qubit to temperatures arbitrarily close to
absolute zero. Note, however, that for cooling a qubit toward zero,
since $T_{c}/T_{h}<1$, while for positive temperatures, Eq. \eqref{eq: 5}, we need to resort
only to the ratio $E_{3}/E_{1}\rightarrow\infty$,
for negative temperatures we can have $T_{c}>\left|T_{h}\right|$.
Thus, it is possible to take advantage of negative temperatures by
choosing $T_{c}/\left|T_{h}\right|\rightarrow\infty$.

Regarding the ratio $E_{3}/E_{1}$, there is an important remark: positive temperatures require the condition $E_{3}=E_{2}-E_{1}$, which limits the level structures of the qubits needed for AR to work. On the other hand, negative temperatures allow arbitrary configurations, such as $E_{3}=E_{2}=E_{1}$, see Fig. \ref{fig: 1}(b). Besides, in the perfect insulation condition, if we take for example $E_{3}/E_{1}=1$, Eq.\eqref{eq: 5} tells us that it is impossible, starting from finite and positive temperatures, to cool qubit 1 arbitrarily close to absolute zero. Indeed, according to Eq.\eqref{eq: 5}, the smallest possible temperature for $E_{3}/E_{1}=1$ is obtained when $T_c/T_h$ is very small, such that the final temperature of qubit 1 is $T_{1}=T_{c}/2$. 
Nevertheless, for negative temperatures and in the condition of perfect insulation where Eq.\eqref{eq: 6} applies, even if we let $E_{3}/E_{1}=1$ we can still make the ratio $T_{c}/T_{h}$ very large, thus cooling qubit 1 as close to absolute zero as we want.

\section{Conclusion}

In this work we consider the autonomous quantum refrigerator making
use of three qubits, one qubit to be cooled and the other two playing
the role analogous to the spiral and the engine in a classical refrigerator,
as proposed in Ref. \citep{Linden2010}. Making a comparative study
of the functioning of this refrigerator in environments with either
positive temperature or negative temperatures, we show that under
the same conditions, that is, with the same parameters that characterize
the autonomous quantum refrigerator, the use of negative temperature
brings advantages in relation to the use restricted to positive temperatures.
According to our numerical simulations, negative temperatures allow
a greater cooling range, surpassing the positive temperature operation
by about 20\% for some parameters, see Fig. \ref{fig: 4}(a). Furthermore,
while restricting the environments at positive temperatures, the steady-state of qubit 1, which ends cooled at temperature $T_{1}$, depends
on the temperature $T_{h}$ of the hot reservoir, see Fig. \ref{fig: 2}(a),
by using the hot environment at negative temperature the cooling of
qubit 1 occurs for all $T_{h}$, see Fig. \ref{fig: 3}(a). Also,
by imposing the condition of perfect insulation, that is, isolating
the qubit 1 to be cooled from its environment and letting it interact
directly with the other qubits, we show that, as in the case of positive
temperatures, the qubit 1 can be cooled to a temperature $T_{1}$
arbitrarily close to absolute zero. However, even for perfect insulation, it is still possible to demonstrate an advantage of using negative
temperatures. In fact, for $T_{h}>0$, it is necessary to greatly
increase the ratio between the energy $E_{3}$ of the third qubit
and the energy $E_{1}$ of the qubit 1 to be cooled to bring its temperature
$T_{1}$ to values close to absolute zero. On the other hand, using
$T_{h}<0$ is possible to set the ratio $E_{3}/E_{1}$ to reasonably
small values but let the ratio $T_{c}/\left|T_{h}\right|$ very large,
such that the temperature of qubit 1 to be cooled approaches absolute
zero. To summarize, using $T_{h}<0$ to cool a qubit is more powerful
than $T_{h}>0$.

\begin{acknowledgments}
We acknowledge financial support from the Brazilian agencies: CAPES,
financial code 001, CNPq, FAPEG, and FAPESP. This work was performed
as part of the Brazilian National Institute of Science and Technology
(INCT) for Quantum Information Grant No. 465469/2014-0. 
\end{acknowledgments}

\bibliographystyle{apsrev4-1}
\bibliography{References.bib}

\end{document}